\documentclass[12pt]{article}
\usepackage{amsfonts}
\newtheorem{theorem}{Theorem}

\begin{document}

\title{Universal families and quantum control in infinite dimensions}
\author{R. Vilela Mendes\thanks{%
IPFN - EURATOM/IST Association, Instituto Superior T\'{e}cnico, Av. Rovisco
Pais 1, 1049-001 Lisboa, Portugal, http://label2.ist.utl.pt/vilela/} \thanks{%
CMAF, Complexo Interdisciplinar, Universidade de Lisboa, Av. Gama Pinto, 2 -
1649-003 Lisboa, Portugal, e-mail: vilela@cii.fc.ul.pt}}
\date{ }
\maketitle

\begin{abstract}
In a topological space, a family of continuous mappings is called universal
if its action, in at least one element of the space, is dense. If the
mappings are unitary or trace-preserving completely positive, the notion of
universality is closely related to the notion of controllability in either
closed or open quantum systems. Quantum controllability in infinite
dimensions is discussed in this setting and minimal generators are found for
full control universal families. Some of the requirements of the operators
needed for control in infinite dimensions follow from the properties of the
infinite unitary group. Hence, a brief discussed of this group and their
appropriate mathematical spaces is also included.
\end{abstract}

\section{Quantum control and the infinite-dimensional unitary group.
Essentially infinite-dimensional transformations}

To control the time evolution of quantum systems is an essential step in
many applications of quantum theory\cite{Rabitz1}. Among the fields
requiring accurate control of quantum mechanical evolution are quantum state
engineering, cooling of molecular degrees of freedom, selective excitation,
chemical reactions and quantum computing. A fairly complete characterization
of quantum controllability in finite dimensional spaces is now available.
However, in many situations, for example when continuous spectrum scattering
states are involved\cite{Tarn1}, one has to deal with an
infinite-dimensional Hilbert space.

Wu, Tarn and Li \cite{Tarn2} (see also \cite{Clark1},\cite{Bloch})
established controllability criteria on the infinite-dimensional manifolds
that are generated by non-compact Lie algebras. However, left open is the
question of when these manifolds are dense on the Hilbert sphere, which
would the key requirement for complete controllability in infinite
dimensions.

Dealing with controllability in infinite dimensions one is faced at the
start with the choice of the proper spaces where the (in general unbounded)
control operators are going to act. A good starting point is to find a
suitable mathematical setting for the groups $U\left( \infty \right) $ or $%
O\left( \infty \right) $, which are clearly transitive in infinite
dimensions. Consider a Gelfand triplet 
\[
S^{*}\supset L^{2}\left( \Bbb{R}^{d}\right) \supset S 
\]
$S$ being a nuclear space obtained as the limit of a sequence of Hilbert
spaces with successively larger norms. An element $g$ of $U\left( \infty
\right) $ is a transformation in $S$ such that 
\[
\left\| g\xi \right\| =\left\| \xi \right\| 
\]
By duality $\left\langle x,g\xi \right\rangle =\left\langle g^{*}x,\xi
\right\rangle $, $x\in S^{*},\xi \in S$, the infinite-dimensional unitary
group is also defined on $S^{*}$, the two groups being algebraically
isomorphic.

For the harmonic analysis on $U\left( \infty \right) $ one needs functionals
on $S^{*}$. $U\left( \infty \right) $ is a complexification of $O\left(
\infty \right) $, the infinite-dimensional orthogonal group and a standard
result states that if a measure $\mu $ is invariant under $O\left( \infty
\right) $ it must be of the form 
\[
\mu =a\delta _{0}+\int \mu _{\sigma }dm\left( \sigma \right) 
\]
a sum of a delta and Gaussian measures $\mu _{\sigma }$ with variance $%
\sigma ^{2}$. Hence we are led to consider the $\left( L^{2}\right) $ space
of functionals on $S^{*}$ with a $O\left( \infty \right) -$invariant
Gaussian measure 
\[
\left( L^{2}\right) =L^{2}(S^{*},B,\mu ) 
\]
$B$ being generated by the cylinder sets in $S^{*}$ and $\mu $ the measure
with characteristic functional 
\[
C\left( f\right) =\int_{S^{*}}e^{i\left\langle x,f\right\rangle }d\mu \left(
x\right) =e^{-\frac{1}{2}\left\| f\right\| ^{2}},\hspace{1cm}x\in S^{*},f\in
S 
\]

In conclusion: the proper framework to study transitive actions and
functional analysis in infinite dimensional quantum spaces is the complex
white noise setting\cite{Hida1}. In this context many useful results are
already available. For example, the regular representation of $U\left(
\infty \right) $%
\[
U_{g}\varphi \left( z\right) =\varphi \left( g^{*}z\right) ,\hspace{1cm}z\in
S_{c}^{*},\varphi \in \left( L_{c}^{2}\right) \cong \left( L^{2}\right)
\otimes \left( L^{2}\right) 
\]
splits into irreducible representations\cite{Okamoto} corresponding to the
Fock space (chaos expansion) decomposition of $\left( L_{c}^{2}\right) $%
\[
\left( L^{2}\right) =\oplus _{n=0}^{\infty }\left( \oplus
_{k=0}^{n}H_{n-k,k}\right) 
\]
$H_{n-k,k}$ being a complex Fourier-Hermite polynomial of degree $\left(
n-k\right) $ in $\left\langle z,\xi \right\rangle $ and of degree $k$ in $%
\left\langle \stackrel{\_}{z},\stackrel{\_}{\xi }\right\rangle $

Furthermore, some results concerning a classification of the subgroups of $%
U\left( \infty \right) $ are useful for our purposes. In particular one must
distinguish between subgroups that only involve transformations that may be
approximated by finite-dimensional transformations like $G_{\infty }$,
obtained as the limit of a sequence of finite-dimensional unitary groups 
\[
G_{n}=\left\{ g\in U\left( \infty \right) ,\left. g\right| _{V_{n}}\in
U\left( n\right) ,\left. g\right| _{V_{n}^{\bot }}=I\right\} 
\]
\[
G_{\infty }=\textnormal{proj.}\lim_{n\rightarrow \infty }G_{n} 
\]
from those that contain transformations changing, in a significant way,
infinitely many coordinates. These group elements are called \textit{%
essentially infinite-dimensional} (see \cite{Hida1} for a rigorous
definition and examples). The essential point to retain for our purposes is
that to generate $U\left( \infty \right) $, and therefore to be transitive
in infinite dimensions, some essentially infinite dimensional elements are
needed. The results in the following sections show that one such
transformation is enough.

\section{Unitary control in infinite-dimensions. Universal families}

Given a topological space $X$ and a family of continuous mappings $T_{\alpha
}:X\rightarrow X$ with $\alpha $ belonging to some index set $I$, an element 
$x\in X$ is called \textit{universal} if the set 
\[
\left\{ T_{\alpha }x:\alpha \in I\right\} 
\]
is dense in $X$. The family $\left\{ T_{\alpha }:\alpha \in I\right\} $ will
be called universal if there is at least one universal element $x\in X$.
The problem of quantum controllability for closed systems is therefore the
search for universal unitary families in the Hilbert sphere with a dense set
of universal vectors.

A particularly interesting situation occur when the universal family is
generated by a single operator. If the universal family consists of the
powers $T^{n}$ of a single operators, this one is called \textit{hypercyclic}
and if it is 
\[
\left\{ \lambda T^{n}x\right\} 
\]
with $\lambda $ a scalar, that is dense in $X$, the operator is called 
\textit{supercyclic}. Because all these notions are related to the density
of a set, they depend on the topology of $X$. An interesting fact is that
hypercyclicity is a purely infinite-dimensional phenomenon. No linear
operator on a finite-dimensional space is hypercyclic, as can easily be seen
by considering the operator in its Jordan normal form\cite{Rolewicz}.

A universal unitary family in the infinite dimensional Hilbert sphere has
been found in \cite{Witold}. Because this result will later be generalized
for open systems, I recall here the relevant definitions. By the choice of a
countable basis any separable Hilbert space is shown to be isomorphic to $%
\ell ^{2}\left( \Bbb{Z}\right) $, the space of double-infinite
square-integrable sequences 
\[
a=\left\{ \cdots ,a_{-2},a_{-1},a_{0},a_{1},a_{2},\cdots \right\} \in \ell
^{2}\left( \Bbb{Z}\right) 
\]
\[
\left| a\right| =\left( \sum_{-\infty }^{\infty }\left| a_{k}\right|
^{2}\right) ^{\frac{1}{2}}<\infty 
\]
with basis 
\[
e_{k}=\left\{ \cdots ,0,0,1_{k},0,0,\cdots \right\} 
\]
The following operators are defined:

(i) A linear operator $T_{+}$ acting as a shift on the basis states 
\[
T_{+}e_{k}=e_{k+1},\qquad k\in \Bbb{Z} 
\]
and its inverse 
\[
T_{+}^{-1}e_{k}=e_{k-1},\qquad k\in \Bbb{Z} 
\]

(ii) A $U\left( 2\right) $ group operating in the linear space spanned by $%
e_{0}$ and $e_{1}$ and leaving the complementary space unchanged.

Let $G\left( T_{+},U\left( 2\right) \right) $ be the group generated by
these operators. Then,

\begin{theorem}
\cite{Witold} For any $a\in \ell ^{2}\left( \Bbb{Z}\right) $,$\left|
a\right| =1$, $G\left( T_{+},U\left( 2\right) \right) a$ is dense in the $%
\ell ^{2}\left( \Bbb{Z}\right) -$Hilbert sphere.
\end{theorem}

That is, $T_{+}$ and $U\left( 2\right) $ generate a universal family in the
Hilbert sphere with a dense set of universal vectors.

$T_{+}$ and $U\left( 2\right) $ is already a relatively small set of
generators, but an interesting question is whether a smaller set may be
found, namely whether there are unitary hypercyclic or supercyclic
operators. The answer depends both on the topology of the space and on the
nature of the measure $\mu $ used for the $L^{2}\left( \mu \right) $ space.
With the norm topology in the space $X$, the answer is negative because no
hyponormal operator $\left( \left\| Tx\right\| \geq \left\| T^{*}x\right\|
;x\in X\right) $ can be hypercyclic\cite{Kitai} or supercyclic\cite{Bourdon}.

The situation is different if density in the space $X$ is relative to the
weak topology, with neighborhood basis 
\[
N\left( \psi _{1}\cdots \psi _{n},\varepsilon _{1}\cdots \varepsilon
_{n}\right) =\left\{ \phi :\left| \left\langle \psi _{i}|\phi \right\rangle
\right| <\varepsilon _{i}\right\} 
\]
Then there are weakly supercyclic normal operators which are necessarily
multiples of unitary operators and an example of a unitary hypercyclic
operator has been constructed in a $L^{2}\left( \mu \right) $ space\cite
{Bayart1}. This construction is somewhat particular in that $\mu $ is a
singular continuous measure in a thin Kronecker set. For measures that are
absolutely continuous with respect to the Lebesgue measure one has no weakly
supercyclic operator. Nevertheless a set is usually considered as ``large''
if it carries a probability measure $\mu $ for which the Fourier
coefficients $\stackrel{\symbol{94}}{\mu }\left( n\right) $ vanish at
infinity. It has recently been proved that there is such a probability
measure for which the corresponding $L^{2}\left( \mu \right) $ space has a
weakly supercyclic operator\cite{Shkarin}.

These results raise the interesting possibility that in some quantum spaces
associated to singular continuous measures (hierarchical systems, for
example), complete infinite-dimensional quantum controllability might be
implemented with a single operator and its powers.

\section{A universal family for Kraus operators}

For open systems I will restrict myself to evolutions by completely positive
trace-preserving maps $\Phi $, which may be represented by the Kraus
operator sum representation 
\[
\Phi \left( \rho \right) =\sum K_{i}\rho K_{i}^{\dagger } 
\]
The problem of quantum control in this setting corresponds to the search for
a universal family of operators acting in the operator algebra of bounded
operators $B\left( H\right) $ in the infinite-dimensional Hilbert space $H$.
No countable subset of $B\left( H\right) $ can be dense in the operator norm
topology. Therefore, because one is always interested in control by a
sequence of transformations, the problem has no practical sense in this
topology. Instead one should discuss density in the strong operator
topology, that is, the one with neighborhood basis 
\[
N\left( x_{i},\varepsilon _{i};i=1\cdots n\right) =\left\{ O:\left\|
Ox_{i}\right\| <\varepsilon _{i}\right\} 
\]
The $B\left( H\right) $ operator algebra is separable in this topology,
meaning that any element may be approximated arbitrarily close by some $%
n\times n$ matrix.

Wu, Pechen, Brif and Rabitz\cite{Rabitz2} established general controllability
conditions for Kraus operators. Here, as before, one looks for a minimal set
of generators of a universal family that insures controllability in infinite
dimensions. Consider a separable Hilbert space isomorphic to $\ell
^{2}\left( \Bbb{Z}\right) $, the shift operator $T_{+}$ and its inverse $%
T_{+}^{-1}$, as well as a $U\left( 2\right) $ group acting on the subspace $%
\left\{ e_{0},e_{1}\right\} $ and leaving the complementary space unchanged.
From these one constructs the following useful operators:

- An operator $\Pi $ that exchanges the basis vectors $e_{0}$ and $e_{1}$
and keeps the others unchanged. It is an element of the $U\left( 2\right) $
group,

- The operator $\Pi _{n}=T_{+}^{n}\Pi T_{+}^{-n}$ that exchanges $e_{n}$
with $e_{n+1}$ and keeps the others unchanged,

- The operators $\Pi _{k,k+p}=\Pi _{k}\Pi _{k+1}\cdots \Pi _{k+p-2}\Pi
_{k+p-1}\cdots \Pi _{k+1}\Pi _{k}$ that exchange $e_{k}$ and $e_{k+p}$,
keeping the others unchanged.

As one may expect from Theorem 1, this set of operators, generating all
unitaries in arbitrary dimensions, may also be able to generate all
random-unitary transformations (Kraus operators proportional to unitaries)
but not all trace-preserving completely positive operations. Hence a new
operator must be added, which I will choose to be the projection on a basis
state, for example $P_{0}=\left| e_{0}\right\rangle \left\langle
e_{0}\right| $.

\begin{theorem}
$P_{0},T_{+},T_{+}^{-1}$ and $U\left( 2\right) $ generate a (strong operator
topology-) universal family in the set of all density operators in infinite
dimensions, with a dense set of universal elements.
\end{theorem}

\textbf{Proof:}

Let $\rho $ be an arbitrary density operator in a $n-$dimensional subspace $%
V_{n}$. First, by using the shift operators $T_{+}$ and $T_{+}^{-1}$ , one
can translate the $V_{n}$ subspace in such a way that it contains the basis
vectors $e_{0}$ and $e_{1}$. By the construction in the proof of Lemma 2 of
ref.\cite{Witold} one knows that any normalized vector in an $n-$dimensional
subspace may be transformed by $T_{+},T_{+}^{-1}$ and $U\left( 2\right) $ to
an arbitrary basis state (say $e_{0}$) in the $n-$dimensional subspace. That
means that from $T_{+},T_{+}^{-1}$ and $U\left( 2\right) $ one generates all 
$U\left( n\right) $ transformations. Therefore, with these transformations $%
\rho $ may be brought to its diagonal form $\rho _{D}$. Now to $\rho _{D}$
one applies the Kraus transformation 
\[
\sum_{i=1}^{n}K_{i}\rho _{D}K_{i}^{\dagger }
\]
the Kraus operators being $K_{i}=P_{0}\Pi _{0,i}$ ($i=0,\cdots ,n-1$) ($\Pi
_{0,0}$ is just the identity, an element of the $U\left( 2\right) $ group).
This transforms $\rho _{D}$ into the single projector $P_{0}=\left|
e_{0}\right\rangle \left\langle e_{0}\right| $.

Conversely by applying the Kraus operators $K_{i}=\sqrt{\rho _{D,i}}\Pi
_{0,i}$ to $P_{0}$ and reversing the operations of the unitary group and the
shift, $P_{0}$ may be transformed into any density operator of any other $m-$%
dimensional subspace.

The fact that the density operators in finite-dimensional subspaces are
dense (in the strong operator topology) on the set of all the density
operators in infinite dimensions, completes the proof.

\section{Remarks and conclusions}

1) A relatively small set of operations is sufficient to insure
controllability in infinite dimensions, both for closed and open quantum
systems. The results were established for a general separable Hilbert space.
An even smaller set of operations might be possible for some $L^{2}\left(
\mu \right) $ spaces with singular continuous measures.

2) An essential point that follows from the structure of the infinite
dimensional unitary group is that, to cover densely the Hilbert sphere, 
essentially infinite dimensional elements are needed, that is, transformations that
change significantly an infinite number of components in some basis. One
sees that one such transformation is sufficient in the generating set of the
universal family. Here that role is played by the shift operator which may
have a simple physical interpretation as, for example, the application of a
magnetic field pulse to a system behaving like a charged plane rotator.
Depending of the concrete physical system to be controlled, an appropriate
set of controlling operators should be chosen. In any case, the message to
retain is that full controllability in infinite dimensions requires
essentially infinite dimensional transformations.

3) To the minimal set used for unitary control, an additional generator must
be added to obtain an universal family for Kraus controllability. Here, a
projection operator was used for this purpose. The essential role of one
such transformation suggests that the control scheme of
measurement-plus-evolution\cite{Vilela1} \cite{Clark2} \cite{Rabitz3}, a
simple matter of convenience to extend the set of controllable closed
systems,  is indeed a most natural one for open systems.

\end{document}